\documentclass[doublecol]{epl2}
\usepackage{graphicx}
\usepackage{amssymb}
\usepackage{bm}
\usepackage{color}
\usepackage{multirow}

\begin{document}

\title{Spin-orbital physics for $p$ orbitals in alkali $R$O$_2$
       hyperoxides\\
       --- Generalization of the Goodenough-Kanamori rules}

\shorttitle{Spin-orbital physics for $p$ orbitals in alkali $R$O$_2$
            hyperoxides}

\author {   Krzysztof Wohlfeld\inst{1} \and Maria Daghofer\inst{1} \and
            Andrzej M. Ole{\'s}\inst{2,3}  }

\shortauthor{ K. Wohlfeld, M. Daghofer and A. M. Ole\'{s} }

\institute{
  \inst{1} IFW Dresden, P. O. Box 27 01 16, D-01171 Dresden,
           Germany\\
  \inst{2} Max-Planck-Institut f\"ur Festk\"orperforschung,
           Heisenbergstrasse 1, D-70569 Stuttgart, Germany \\
  \inst{3} Marian Smoluchowski Institute of Physics, Jagellonian
           University, Reymonta 4, PL-30059 Krak\'ow, Poland
}

\date{\today}

\pacs{75.25.Dk}{Orbital, charge, and other orders,
                including coupling of these orders }

\abstract{ We derive a realistic spin-orbital model at finite
Hund's exchange for alkali hyperoxides. 
We find that, due to the geometric frustration of the oxygen lattice
spin and orbital waves destabilize both spin and $p$-orbital order in almost all potential 
ground states. We show that the orbital order induced by
the lattice overrules the one favoured by superexchange and that
this, together with the large interorbital hopping, leads to
generalized Goodenough-Kanamori rules. They ($i$) lift the
geometric frustration of the lattice, and ($ii$) explain the
observed layered $C$-type antiferromagnetic order in alkali
hyperoxides. This is confirmed by a~spin-wave dispersion with no
soft-mode behavior presented here as a prediction for future
experiments. }

\maketitle

Alkali $R$O$_2$ (with $R$=K,Rb,Cs) hyperoxides attracted a lot of
attention in the 70s and 80s \cite{Lab79} but then have been
overshadowed by various classes of transition metal oxides ---
largely due to the discovery of the high temperature superconductivity
and colossal magnetoresistance in the latter. These and other
fascinating phenomena arise in transition metal oxides due to strong
local Coulomb correlations within party filled $d$ orbitals
\cite{Ima98}. A particular class of these compounds are systems with
orbital degeneracy in which the effective low-energy interactions
involve not only spin but also orbital degrees of freedom within the
spin-orbital superexchange \cite{Kug82,Nag00}. One of its consequences
are rather complex phase diagrams in doped manganites \cite{Tok06} that
follow from competing magnetic interactions in unfrustrated perovskite
lattice. These systems are of great interest at present because orbital
superexchange interactions are directional and thus intrinsically
frustrated \cite{Fei97}. Following this idea, purely orbital
frustrated models were developed and serve as paradigmatic models for
investigating order-disorder phenomena and quantum phase transitions
\cite{Nus04,Nus08,Tro10}. On one hand, such interactions are usually
inherently coupled to spin interactions and such exotic phenomena as
joint spin-orbital excitations \cite{Fei97} or entangled states
\cite{Ole06} arise. On the other hand, they also couple to lattice
distortions that may remove frustration and stabilize magnetic order
\cite{Zaa93,Mot99,vdB01,Gru02,Zho06}.

Quite recently, it was realized that spin-orbital physics with 
$p$ orbitals determines the physical properties of $R$O$_2$
hyperoxides \cite{Sol08, Kov09, Kum10, Ylv10}. While an
independent-electron picture suggests that the $R$O$_2$
hyperoxides are FM halfmetals, they are in fact Mott insulators
with one hole shared between the two antibonding O$_2$ molecular
$p$ orbitals \cite{Sol08}. Thus the localized hole has an orbital 
$p$ degree of freedom (in addition to spin). Along with solid oxygen
\cite{Mei84}, the alkali hyperoxides constitute one of the few
examples of defect-free $p$-band Mott insulators in condensed
matter systems \cite{Sol08, Kov09}: thus they share certain common
features of the above mentioned transition metal oxides with $3d$
electrons, and with the novel $p$-orbital systems in optical lattices
\cite{Lew11, Wir11, Wu07, Zha08}. Yet, despite the recent interest
in these compounds, a central question concerning their properties
has not been answered: what is the origin of the same magnetic
order observed in the $R$O$_2$ hyperoxides below a N\'eel
temperature that varies between 5 to 15 K\cite{Lab79}? As we show
below, the antiferromagnetic (AF) order is here indeed due to a
different mechanism than the ones usually discussed in
transition metal oxides \cite{Ima98}, namely a frustration between
lattice-driven and correlation-driven effects.

\begin{figure}[t!]
\begin{center}
   \includegraphics[width=4.0cm]{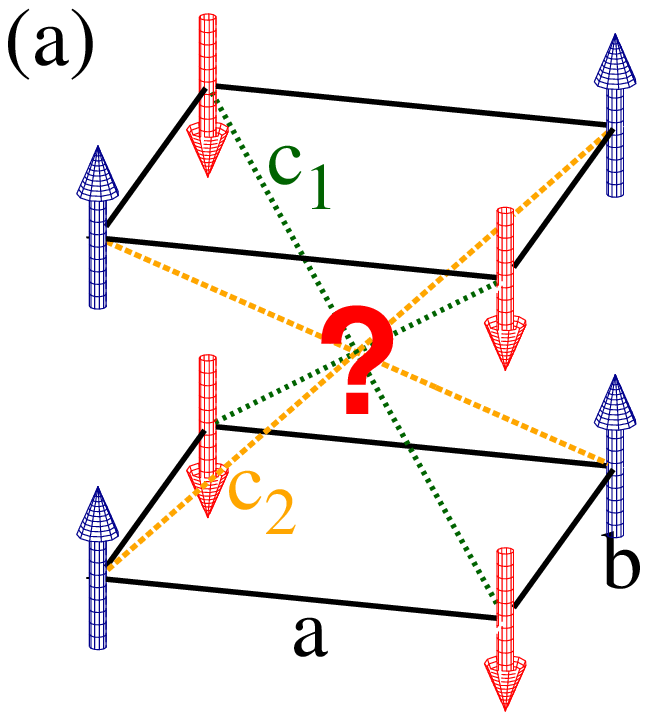} \quad
   \includegraphics[width=3.85cm]{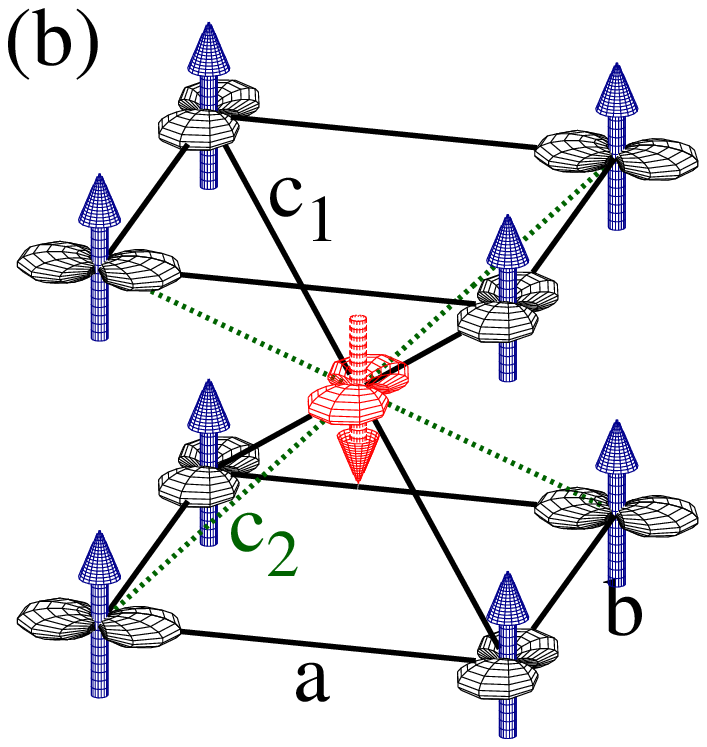}
\end{center}
\caption{(Colour on-line)
(a): Spin Heisenberg model on the bct lattice
is frustrated. (b): Yet, a layered
$C$-AF spin order
is stable in alkali $R$O$_2$ hyperoxides \cite{Lab79}.
Orbital ordering might explain it, though a violation of
Goodenough-Kanamori rules is inevitable in 
at least one plane (here shown along the dotted $c_2$ bonds), 
see text. } 
\label{fig:1}
\end{figure}

A `perfect' AF order, with opposite spins along all
nearest-neighbour bonds, is excluded in the frustrated body
centered tetragonal  (bct) lattice common for all alkali
hyperoxides, see Fig. \ref{fig:1}(a). The observed magnetic order
is instead a layered $C$-type antiferromagnetic ($C$-AF) order,
with ferromagnetic (FM) $ab$ planes and AF $c_1$ and $c_2$ bonds,
shown in Fig. \ref{fig:1}(b). As we discuss in more detail below,
the geometric frustration continues to play here a fundamental
role via the Goodenough-Kanamori rules (GKR) \cite{Goo63} and
tends to destabilize the $C$-AF order as well. It turns out that
the observed magnetic order can only arise when the
well-established (classical) GKR are not obeyed and they are
replaced by {\it generalized} GKR. We show that the competition
between superexchange and a generic Jahn-Teller (JT) effect leads
to an orbital pattern with substantial interorbital hopping and
that this in turn induces the generalized GKR driving the $C$-type
AF order. The experimentally observed magnetic order thus arises
from rather subtle interplay between spin-orbital physics and
orbital-lattice coupling present in the alkali hyperoxides, being
strikingly different from both $d$-orbital physics in
transition metal oxides, and also from the $p$-orbital optical
lattices.

First we show that geometric frustration is incompatible with the
GKR , which state \cite{Goo63} that a bond with alternating
orbital (AO) order leads to FM spin exchange, while ferro-orbital
(FO) order induces AF spin coupling. A $C$-type AF order can thus
arise if bonds within the FM $ab$ plane show AO order, and those
in the AF $c$ directions should show FO order. In many
transition metal oxides, such coexisting spin-orbital order
arises, e.g., in the archetypal orbital system KCuF$_3$
\cite{Kug82}. However, this mechanism is here frustrated due to
the bct lattice geometry, in a similar manner as the `perfectly'
AF state depicted in Fig.~\ref{fig:1}(a) --- the experimentally
found FM order within the planes requires AO order along $a$ and
$b$ bonds. In the next higher plane along the $c$ direction, see
the shaded orbital and spin in Fig.~\ref{fig:1}(b), either choice
of the orbital leads to some bonds with AO order and other bonds
with FO order. Consequently, the GKR would imply FM spin exchange
either along $c_2$ [as in Fig.~\ref{fig:1}(b)] or along $c_1$,
while the observed order is AF along both $c_1$ and $c_2$. The
$C$-type AF order thus violates the GKR due to geometric
frustration on the bct lattice.

Since the magnetic order is AF along $c_2$ bonds, the effective FM
magnetic interaction predicted by the GKR would have the `wrong'
sign. Such frustrated `wrong' couplings can in principle still be
compatible with long-range order --- cf. a $J_1-J_2$ model on a
square lattice \cite{Cha88} or if the GKR violation is driven by
spin-orbital entanglement \cite{Ole06}. But we show below that in
the present case, the `wrong' signs lead to soft modes in magnetic
and orbital excitations which destroy long-range order
--- this phenomenon is somewhat similar to
the collapse of order due to enhanced quantum fluctuations in
coupled spin-orbital systems \cite{Fei97}. The geometric
frustration in the bct lattice thus not only leads to violation of
the GKR, but also destabilizes magnetic order whenever these rules
are violated. We are going to show that the `way out' suggested by
the alkali hyperoxides are the generalized GKR explained below.

In the Mott-insulating limit of strong intraorbital Coulomb
repulsion $U$ applicable to $R$O$_2$~\cite{Sol08,Ylv10}, the
interacting spin and $p$-orbital degrees of freedom can be
described by a spin-orbital Hamiltonian. In an orbital basis given
by the $p_x$ and $p_y$ orbital, one finds the superexchange
Hamiltonian for finite Hund's exchange $\eta\equiv J_H/U$
\begin{equation}\label{eq:h}
\mathcal{H}= \sum_{\langle {\bf i} {\bf j} \rangle || \gamma}
\left\{ \hat{J}^{\gamma}_{{\bf i}{\bf j}} \ ({\bf S}_{\bf i }\cdot
{\bf S}_{\bf j})+\hat{K}^{\gamma}_{{\bf i}{\bf j}}\right\},
\end{equation}
where $\gamma\in\{ab, c_1, c_2\}$ denotes the bond direction
and the orbital operators modulating the magnetic exchange are:
\begin{eqnarray}
\hat{J}^{ab}_{{\bf i} {\bf j}} / J_\sigma&=& \alpha
\left(r_{13} T^x_{\bf i} T^x_{\bf j}
+ r_{12\bar{3}} T^y_{\bf i} T^y_{\bf j}\right)\nonumber \\
&+&\frac{1+\alpha^2}{2} \left(r_{13} T^z_{\bf i} T^z_{\bf j} -
\frac{r_{1\bar{2}\bar{3}}}{4}\right),\\
\hat{K}^{ab}_{{\bf i} {\bf j}} / J_\sigma &=&  \frac14 \alpha
\left({R_{1\bar{3}}}T^x_{\bf i} T^x_{\bf j}
+  R_{1\bar{2}{3}} T^y_{\bf i} T^y_{\bf j} \right)\nonumber \\
& +&\frac{1+\alpha^2}{8} {R_{1\bar{3}}} T^z_{\bf i} T^z_{\bf j},\\
\hat{J}^{c1, c2}_{{\bf i}{\bf j}}\!/ J_{xx}\!\!&=& (1+\beta^2)
\left(r_{13} T^x_{\bf i} T^x_{\bf j}
-\frac{r_{1\bar{2}\bar{3}}}{4} \right)\nonumber \\
&\pm& \beta r_{23} (T^x_{\bf i} + T^x_{\bf j}) \nonumber \\
&+& (1-\beta^2) \left( r_{12\bar{3}} T^y_{\bf i} T^y_{\bf j}
+r_{13} T^z_{\bf i} T^z_{\bf j}  \right),\\
\hat{K}^{c1, c2}_{{\bf i}{\bf j}}\!/ J_{xx}\!\! &=&\! \frac14\,(1+\beta^2)
R_{1\bar{3}} T^x_{\bf i} T^x_{\bf j}
\mp \frac14 \beta {r_{23}}(T^x_{\bf i} + T^x_{\bf j}) \nonumber \\
&+&\!\frac14\,(1-\beta^2) \left( R_{1\bar{2}{3}} T^y_{\bf i} T^y_{\bf j} +
R_{1\bar{3}} T^z_{\bf i} T^z_{\bf j} \right).
\end{eqnarray}
Here ${\bf S}_{\bf i}$ are spin $S=1/2$ operators, and ${\bf
T}_{\bf i}\equiv\{T_{\bf i}^x,T_{\bf i}^x,T_{\bf i}^z\}$ are
$T=1/2$ orbital pseudospin operators for $p$ orbitals $a(b)$ (see
footnote
\footnote{ For simplicity we neglect a small
superexchange term which is different along $a$ and $b$ direction;
we have verified that it does not lead to distinct results. }),
with electron number operators $\{n_{{\bf i}a},n_{{\bf i}b}\}$,
and $T^z_{\bf i}=(n_{{\bf i}a}-n_{{\bf i}b})/2$. Although
interorbital hopping within the $ab$ plane vanishes in the chosen
orbital basis $\{p_x,p_y\}$, different longitudinal ($t_\sigma$)
and transverse ($t_\pi$) hoppings lead to rather involved
superexchange terms: Ising terms $\propto J_\sigma= 4t^2_\sigma/U$
and $\propto\alpha^2 J_\sigma$ and `quantum' terms $\propto\alpha
J_\sigma$, where $\alpha\equiv t_{\pi}/t_{\sigma}$. On the other
hand, between the $ab$ planes (i.e., in the $c_1$ and $c_2$
planes) the diagonal hoppings $t_{xx}$ between each pair of the
same molecular orbitals $aa$ or $bb$ result in superexchange
$\propto J_{xx}=4t^2_{xx}/U$. This is furthermore accompanied by a
substantial interorbital hopping $t_{xy}$ (see below), leading to
additional superexchange channels $\propto \beta J_{xx}$ and
$\propto\beta^2 J_{xx}$, with $\beta\equiv t_{xy}/t_{xx}$ and the
$\pm$ signs corresponding to $(111)$ and $(1\bar{1}1)$ directions.
Hund's exchange contributes via: 
$r_{13}=r_1+r_3$, $r_{23}=r_2+r_3$, $r_{12\bar{3}}=r_1+2r_2-r_3$,
$r_{1\bar{2}\bar{3}}=r_1-2r_2-r_3$, $R_{1\bar{3}}=3r_1-r_3$,
$R_{1\bar{2}3}=3r_1-2r_2+r_3$, where $r_1=1/(1-3\eta)$,
$r_2=1/(1-\eta)$ and $r_3=1/(1+\eta)$.

In what follows we take the units of $J_\sigma\equiv 1$ and assume
a realistic value of $\eta=0.15$ \cite{Sol08, Kum10} and
$J_{xx}/J_\sigma=0.4$ \cite{Ylv10}. We have verified 
that small changes of these two latter parameters (possible
for different $R$O$_2$) do not change the main results of the
paper. However, we vary the transverse hopping $\alpha$ and
interorbital hopping $\beta$, since the phase diagram of the
spin-orbital Hamiltonian Eq. (1) shows quite distinct behaviour
for different parameter regimes. Investigating various regimes
will thus turn out to be illustrative, as different processes are
dominant in each. Based on recent studies which predicted
$(\alpha,\beta) = (0.01,1.90)$ for KO$_2$ \cite{Sol08} and
$(\alpha,\beta) = (0.30,1.77)$ for RbO$_2$ \cite{Ylv10} we suggest
that a {\it realistic} parameter range for these hoppings in
$R$O$_2$ is $\alpha \in [0.0,0.3]$ and $\beta 
\in 
[1.5,2.0]$.

We obtained the classical energies of a large variety of candidate
ground states with at most two sublattices (Monte-Carlo
simulations of the classical model did not indicate larger unit
cells). Since the orbital interactions are not SU(2) symmetric, it
has to be established whether orbital order involves $T^x$ or
$T^z$ pseudospins. 
We have verified that the orbital order of
$T^y$ pseudospins is destabilized by 
orbital waves, similar to the spin-wave case
  discussed in more detail below, for any realistic parameters and it is thus enough to consider
only $T^x$ or $T^z$ orbital order (we omit here a `canted' phase
with pseudospin in the $xz$ plane) 
\cite{vdB99} (see also footnote
\footnote{ The classical Monte Carlo did likewise not indicate any
$T^y$ order. }). The large degeneracy reported in
Ref.~\cite{Ylv10} for $\eta=0$ is partly removed by finite Hund's
exchange $\eta>0$ which splits off the energies of intermediate
$p^2$ states and favours more some superexchange processes.
Still several classical states are very close in energy --- indeed
this feature is generic for frustrated spin-orbital interactions
near orbital degeneracy~\cite{Fei97}. 

Via a mean-field decoupling 
(justified here due to large $\eta$, cf. Ref. \cite{Ole06}),
a given orbital order yields an effective Heisenberg Hamiltonian for 
the spins:
\begin{equation}\label{eq:hs}
 H_{\rm S}\!=
 \!J_{ab}\!\! \sum_{\langle {\bf i} {\bf j} \rangle || ab}\!\!
 {\bf S}_{\bf i }\! \cdot \!{\bf S}_{\bf j}
 +\!J_{c1}\!\! \sum_{\langle {\bf i} {\bf j} \rangle || c1}\!\!
{\bf S}_{\bf i }\! \cdot \!{\bf S}_{\bf j}
 +\!J_{c2}\!\! \sum_{\langle {\bf i} {\bf j} \rangle || c2}\!\!
{\bf S}_{\bf i }\! \cdot \!{\bf S}_{\bf j} ,
\end{equation}
where $\{J_{ab},J_{c1},J_{c2}\}$ are the effective magnetic
exchange constants determined from the spin-orbital model \cite{Ole06},
\begin{equation}
J_\gamma(\phi)\equiv
\langle\phi|\hat{J}^{\gamma}_{{\bf i}{\bf j}}|\phi\rangle,
\label{eq:j}
\end{equation}
$\gamma=ab,c_1,c_2$ and $|\phi\rangle$ is the orbital ground
state. By assuming classical $C$-AF order and determining quantum
corrections via the linear spin-wave theory (LSWT), we now show
that the frustration has a decisive impact on the ground state.

\begin{figure}[t!]
\begin{center}
   \includegraphics[width=4.7cm]{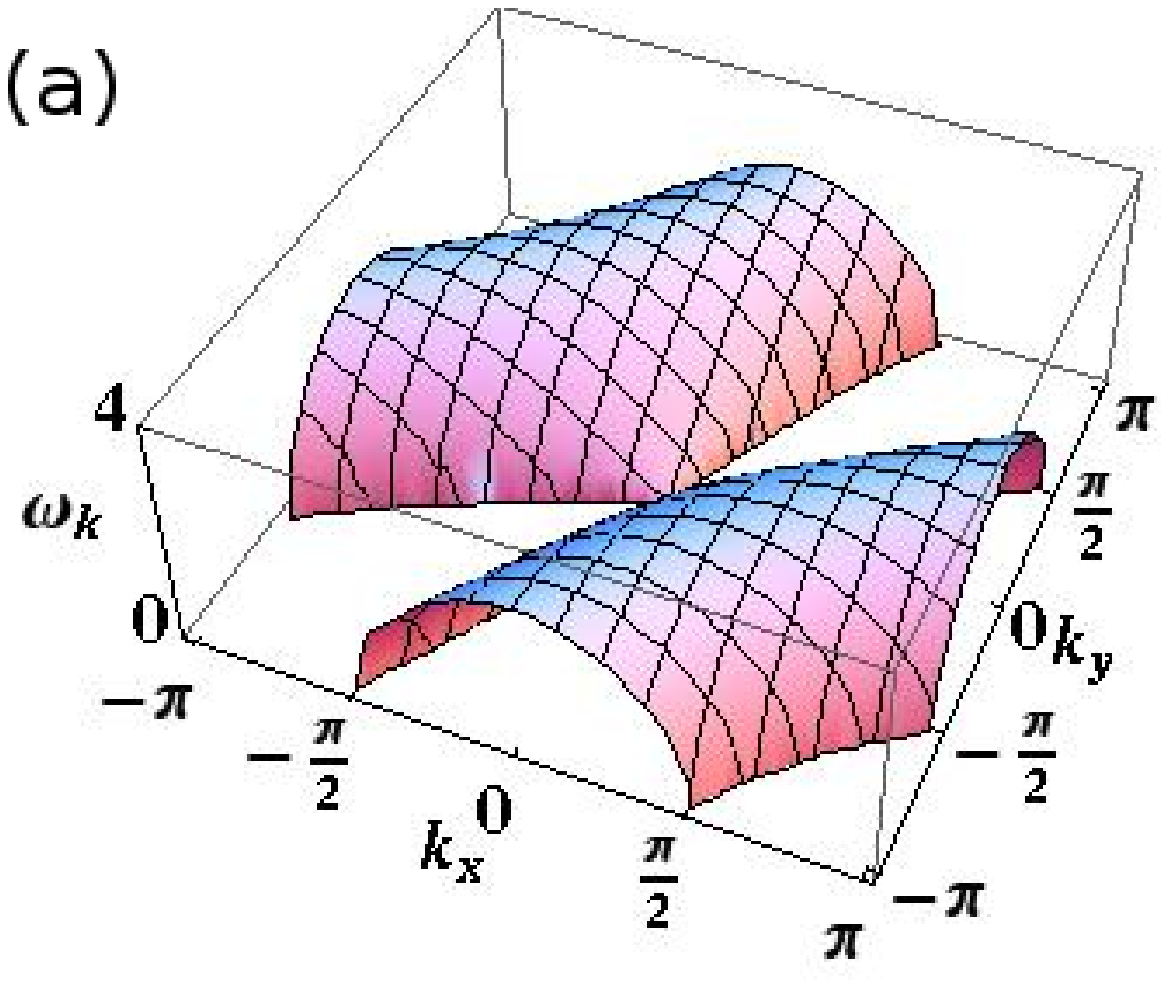}
   \includegraphics[width=3.3cm]{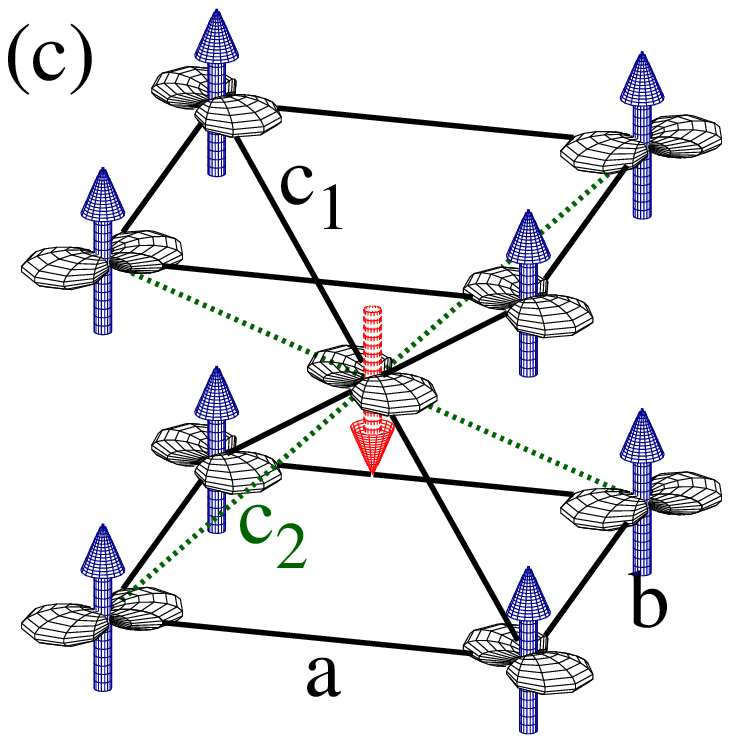} \vspace{10pt} \\
   \includegraphics[width=4.7cm]{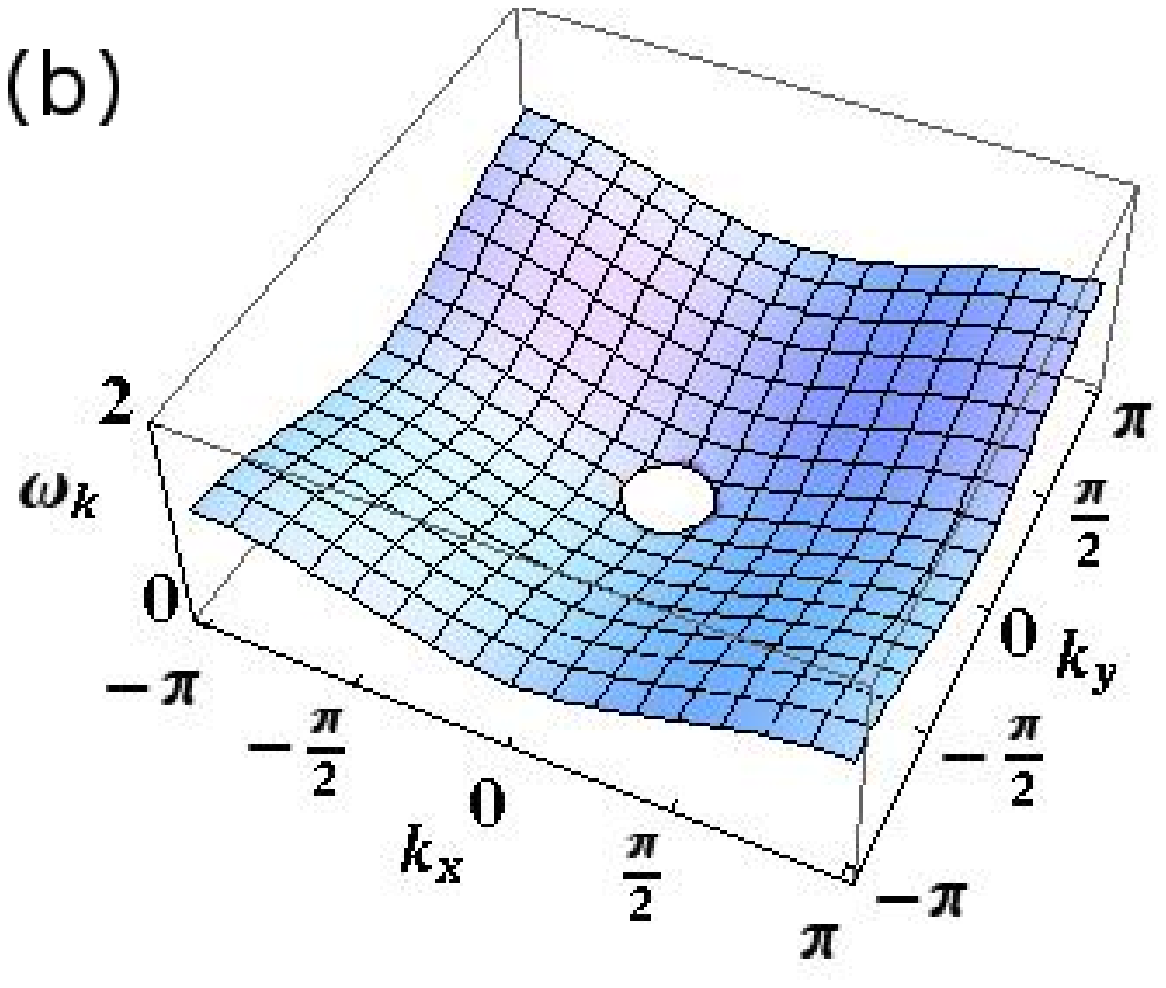}
   \includegraphics[width=3.3cm]{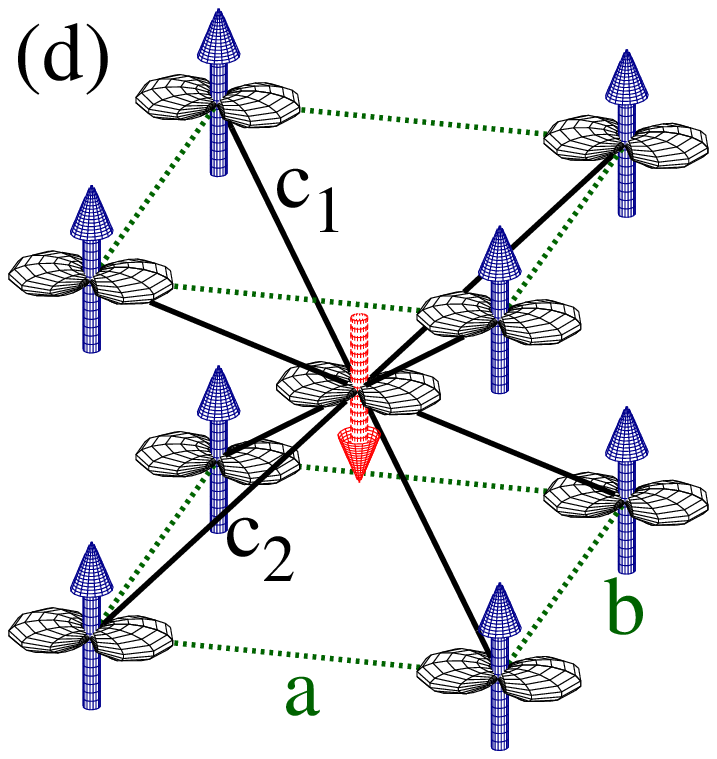}
\end{center}
\caption{(Colour on-line)
Collapse of spin order in the LSWT
[(a,b): soft modes in $\omega_{\bf k}$ at $k_z=0$] for two
representative orbital states: (a,c) --- $C$-AO$x$ order may
explain the FM order in the $ab$ plane according to GKR [solid
bonds in (c)] but spin exchanges in the $c2$ plane have wrong
signs [dashed bonds in (c)]. (b,d) --- FO$z$ order stabilizes the
AF order in the $\{c_1,c_2\}$ planes according to GKR [solid bonds
in (d)], but gives wrong signs of spin exchanges in the $ab$
planes [dashed bonds in (d)].
Parameters: $\alpha=0.30$ and $\beta= 1.77$ \cite{Ylv10}.
}
\label{fig:2}
\end{figure}

Using Holstein-Primakoff bosons $\alpha^\dag_{\bf k}$, after
Fourier and Bogoliubov transformations one obtains from Eq.
(\ref{eq:hs}),
\begin{equation}
H_{\rm S} =\sum_{\bf k}
\omega_{\bf k} \left(\alpha^\dag_{\bf k}\alpha_{\bf k}^{}+\frac12\right),
\end{equation}
with the spin-wave dispersion
\begin{equation} \label{eq:omega}
\omega_{\bf k} = \sqrt{A_{\bf k}^2 - B_{\bf k}^2}\;,
\end{equation}
$A_{\bf k}=2\{J_{ab}(\gamma_{\bf k}-1)+J_{c1}+J_{c2}\}$, and 
$B_{\bf k}=2(J_{c1}\eta_{\bf k}+J_{c2}\zeta_{\bf k})$.
Here $\gamma_{\bf k}=(\cos k_x + \cos k_y)/2$,
$\eta_{\bf k}=\cos(k_x/2 -k_y/2)\cos (k_z/2)$ and
$\zeta_{\bf k}=\cos(k_x/2+k_y/2)\cos (k_z/2)$
follow from the bct lattice structure.
As can be seen in Figs. \ref{fig:2}(a) and \ref{fig:2}(b) 
for two representative orbital states $|\phi\rangle$ (see also below),
$A_{\bf k}^2 < B_{\bf k}^2$ in parts of the Brillouin zone.
This happens when any of the inequalities $J_{ab}<0$, $J_{c1}>0$,
$J_{c2}>0$ is not fulfilled (giving rise to the above mentioned
`wrong' signs of exchange constants) and GKR are violated in at
least one plane
(see also footnote \footnote{There is just one exception to this rule: when the spin exchange 
constant with the `wrong' sign has a smaller absolute value than the other {\it two}
exchange constants with the `correct' sign.
This means that either ($i$) $J_{ab} > 0$, $J_{c1}>0$, 
$J_{c2}>0$, if $|J_{ab}| < |J_{c1}|$
{\it and} $|J_{ab}| < |J_{c2}|$, or ($ii$) $J_{ab} < 0$,
$J_{c1}<0$, $J_{c2}>0$, if $|J_{c1}| < |J_{ab}|$ {\it and}
$|J_{c1}| < |J_{c2}|$ (and similar for $J_{c1}\leftrightarrow
J_{c2}$).}). 
The resulting imaginary energies 
(soft modes) indicate that the ground state is unstable. 
Physically, this is related to the dispersive character of the spin 
waves, i.e., a propagating spin-flip excitation is not balanced by
an ``Ising-like'' local excitation and the ground state collapses.
 
The above can also be seen in the critical case ($\omega_{\bf k}=0$) when 
$A_{\bf k}^2=B_{\bf k}^2$ in parts of the Brillouin zone (this happens if 
$J_{ab} = J_{c1}=J_{c2} > 0$ or $J_{ab} = J_{c1}= - J_{c2} < 0$). Although 
then the energies $\omega_{\bf k}$ remain real, the quantum corrections to 
the order parameter in the harmonic approximation 
$\delta S  = \sum_{\bf k} (A_{\bf k}-\omega_{\bf k})/ (2N \omega_{\bf k})$
diverge 
\footnote{$N$ is the number of lattice sites. When 
$J_{ab} = J_{c1}=J_{c2} > 0$, then $\delta S\sim\int\mathrm{d} k_x \mathrm{d} k_y \mathrm{d} k_z \frac{1}{k_z}$, while when $J_{ab} = J_{c1}= - J_{c2} < 0$, then 
$\delta S\sim\int\mathrm{d} k_x \mathrm{d} k_y \mathrm{d} k_z \{(k_x^2+k_y^2)/(k_x+k_y)\}$.
In both cases $\delta S \rightarrow \infty$.}
due to the onset of soft modes and the classical order is destroyed,
cf. Ref.~\cite{Fei97}.

\begin{figure}[t!]
\begin{center}
   \includegraphics[width=0.24\textwidth]{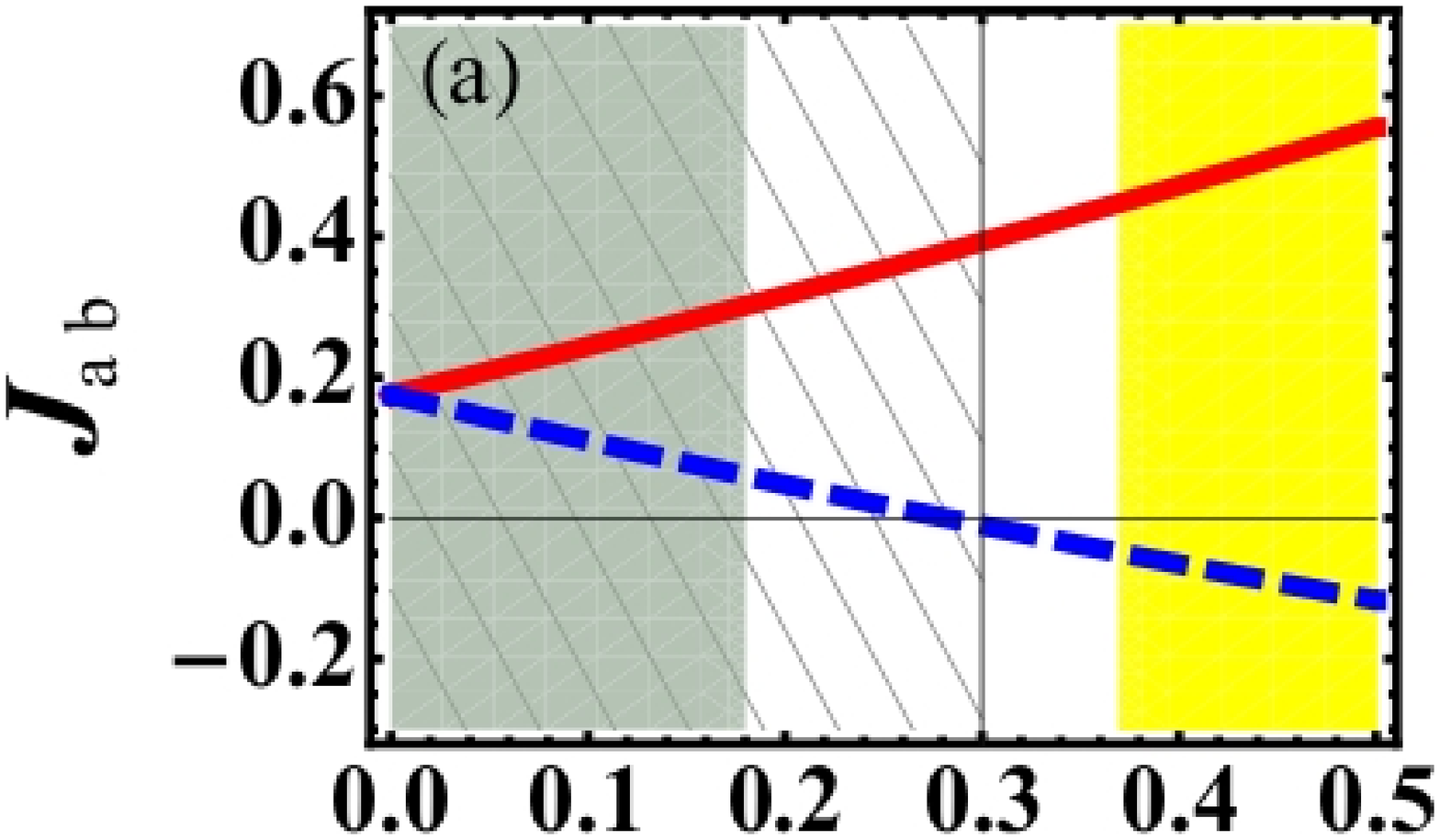}
   \includegraphics[width=0.24\textwidth]{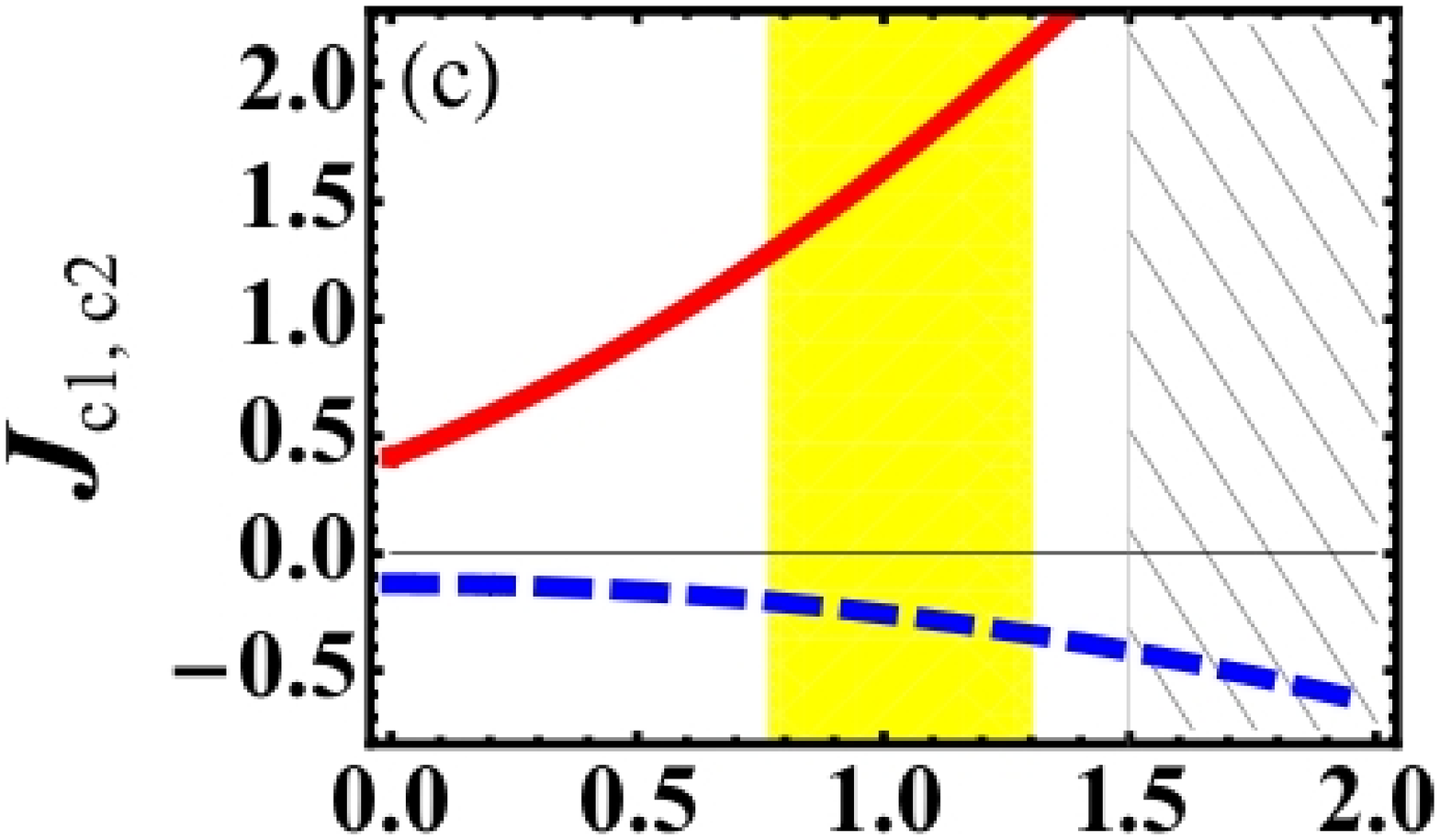} \vskip -.3cm
   \includegraphics[width=0.24\textwidth]{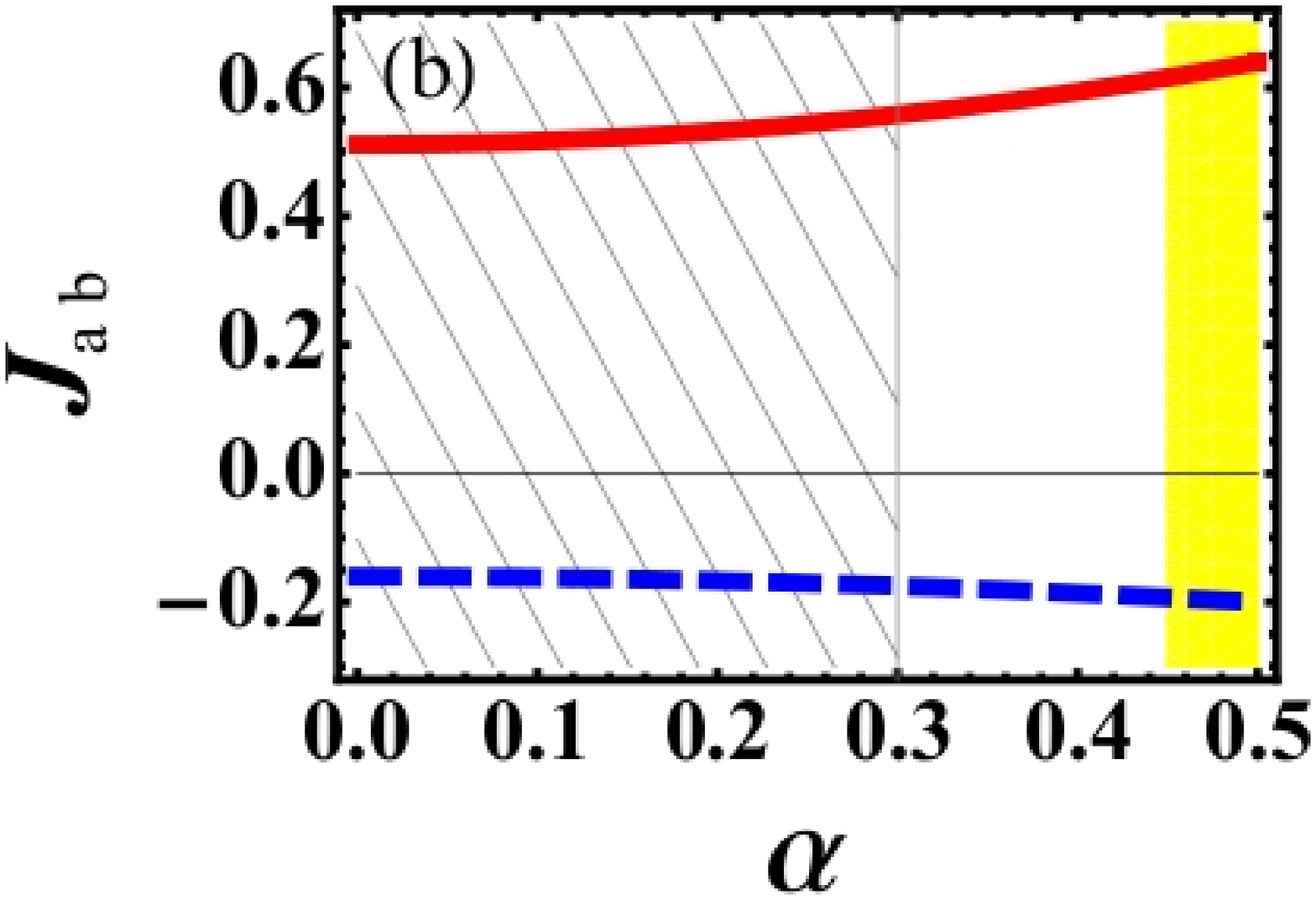}
   \includegraphics[width=0.24\textwidth]{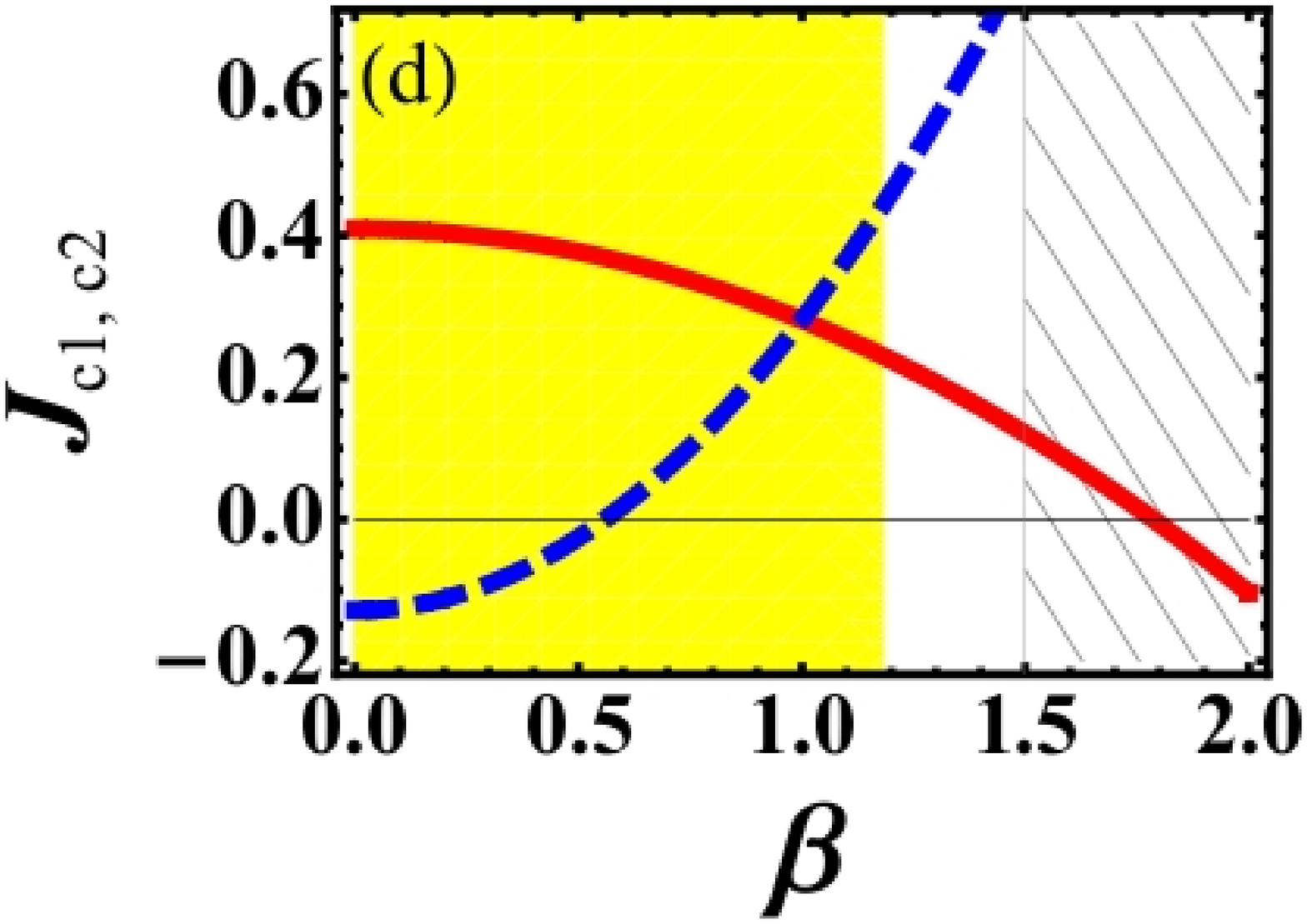}
\end{center}
\caption{(Colour on-line)
Spin exchange constants for different hoppings $\{\alpha,\beta\}$:
(a,b) $J_{ab}$ for increasing $\alpha$, and
(c,d) $J_{c1}$ and $J_{c2}$ for increasing $\beta$;
solid (dashed) lines depict $J_\gamma$ calculated for FO (AO) order.
Panels (a,c) and (b,d) for $T^x$ and $T^z$ pseudospin order.
Areas with oblique lines depict realistic values
values of $\alpha$ and $\beta$ in $R$O$_2$ 
(see also main text).
In shaded areas orbital order is stable for the realistic value 
of $\beta
\in [1.5, 2.0]
$ (left panel) and $\alpha 
\in [0.0,0.3]$ (right panel):
gray (yellow) --- $C$-AO order stable;
dark gray (green) --- FO order stable.
}
\label{fig:3}
\end{figure}

After establishing the importance of having the `correct' sign for
all magnetic exchange constants, we now discuss them more
explicitly for five orbital states: ($i$) $C$-AO$x$
[Fig.~\ref{fig:2}(c)] and FO$x$ (not shown) states with ordered
$T^x$ pseudospin, ($ii$) $C$-AO$z$ [Fig.~\ref{fig:1}(b)] and FO$z$
[Fig.~\ref{fig:2}(d)] states with uniform $T^z$ order, and ($iii$)
an orbital liquid (OL) state with disordered orbitals. 
Note that, as usual \cite{Ima98}, the $C$-AO order means twice 
as many bonds with AO order ($ab$ plane and one of the $c$ planes)
than the FO order (other $c$ plane), though, the choice of FO directions
is different than for $C$-AF order. The latter order generates
a minimal number of
bonds with `wrong signs' in a wide parameter range of $\alpha$ and
$\beta$, see Fig. \ref{fig:3}. We begin with spin exchange
constants for $C$-AO$x$ and FO$z$ phases. Since the GKR are here
perfectly fulfilled for almost all values of $\alpha$ and $\beta$,
one immediately notices that always at least one of the exchange
constants will have a `wrong' sign. Besides, also the magnitudes
of the exchange constants with `wrong' signs are such that for
realistic values of $\beta$ and $\alpha$ (see meshed areas in Fig.
\ref{fig:3}), the $C$-AF cannot be stable for $C$-AO$x$ or FO$z$
order. Finally, even if for some values of $\alpha$ and $\beta$
the GKR are not enforced by the Hamiltonian for these two orbital
states, this only {\it increases} the number of exchange constants
with `wrong' signs. Similarly, also for the OL state (not shown)
the $C$-AF phase is unstable in the entire range of $\eta$.

A different situation, however, arises for the $C$-AO$z$ and FO$x$
states. Here, the $C$-AF phase can be stable for a
wide parameter range of $\alpha$ and $\beta$ (including the values
realized in $R$O$_2$). On one hand, for the FO$x$ case this is
purely due to the fact that the spin exchange with the `wrong'
sign in the $ab$ plane has typically a much smaller magnitude than
the AF one in $c$ planes, see Figs. \ref{fig:3}(a) and
\ref{fig:3}(c). On the other hand, for the $C$-AO$z$ state this is
not only due to the fact that the `wrong' exchange constants have
small magnitude but rather because the GKR are not enforced here
and the signs of the spin exchange constants permit a stable
$C$-AF phase in the LSWT, see Figs. \ref{fig:3}(b) and
\ref{fig:3}(d). More precisely, for a wide range of $\beta$ the
spin exchange constant $J_\gamma$ is positive \emph{both} in the
FO$z$ $c_1$ plane and in the AO$z$ $c_2$ plane (e.g. $J_{\gamma} >
0$ still for FO state with $\beta= 1.77$ as realized in RbO$_2$).
There is thus no frustrated magnetic coupling in the AO$z$ state,
as a lifting of the classical GKR 
permits instead that all spin couplings have the correct sign.

Since the spin order can be so easily destabilized by the spin
excitations,
we have performed a linear orbital-wave theory (cf.
Ref. \cite{vdB99}) and verified that soft modes arise {\it also}
in the orbital wave spectrum. It is remarkable that the only two
orbital states for which the $C$-AF was stable 
with respect to spin
waves (the $C$-AO$z$ and FO$x$ orbital states) collapse now
in a very similar way as shown for the spin case in Fig.
\ref{fig:2} for almost whole range of realistic values of $\alpha$
and $\beta$ in $R$O$_2$, see shaded areas in Fig.~\ref{fig:3}. In
fact, the FO$x$ state can only be stable for small values of $\alpha$ 
which may be realistic for some $R$O$_2$ compounds, but not for
$\alpha=0.3$ as suggested for RbO$_2$. This order
thus cannot explain the origin of the same $C$-AF order stable in
all $R$O$_2$. The physics behind these phenomena is as follows:
($i$) finite transverse hopping $\alpha$ 
enhances orbital fluctuations in the $ab$ plane 
(due to the `wrong' sign of the
orbital interactions for the FO and FM order in the $ab$ plane)
which are for the orbital case of comparable size as in the $c$
planes and this destroys the FO$x$ orbital order; 
($ii$) while the
interorbital hopping $\beta$ (discussed below) turns out to be a
crucial ingredient in stabilizing the observed magnetic order
\emph{provided} the orbital sector shows $C$-AO$z$ order,
precisely this interorbital process suppresses the $C$-AO$z$
order.

\begin{figure}[t!]
\begin{center}
   \includegraphics[width=6.8cm]{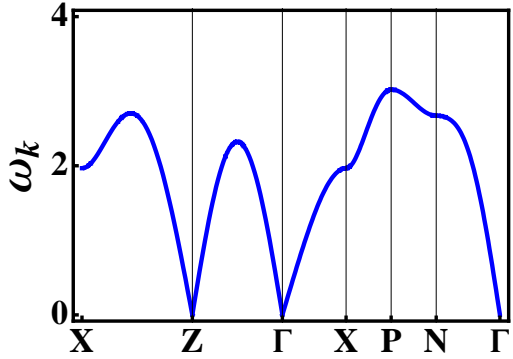}
\end{center}
\vskip -.4cm
\caption{(Colour on-line)
Spin-wave dispersion $\omega_{\bf k}$ along the high symmetry
directions in the Brillouin zone 
for the $C$-AOz phase using Eq. (\ref{eq:omega}).
Parameters: $\alpha=0.30$ and $\beta=1.77$ \cite{Ylv10}.
High symmetry points: $X= (\pi, 0, \pi)$, $Z=(0,0, 2\pi)$,
$\Gamma=(0, 0, 0)$, $P=(\pi, \pi, \pi$), $N=(\pi, 0, \pi)$
cf. Ref. \cite{Sol08} with $a\equiv c \equiv 1$.
}
\label{fig:4}
\end{figure}

However, the orbital order is also sensitive to the
orbital-lattice coupling, stemming from the JT effect, and the
resulting orbital state $|\phi\rangle$ determines spin-wave
dispersion (see below and Fig. \ref{fig:4}). In fact, a standard
and rather weak JT interaction is enough to stabilize $C$-AO$z$
order [cf. Fig. \ref{fig:1}(b)] over the OL phase for realistic values of $\alpha$ and $\beta$. We
have verified that the interaction,
\begin{equation}
\mathcal{H}_{\rm JT}= E_{\rm JT} \sum_{\langle {\bf i} {\bf j}
\rangle || ab} T^z_{\bf i} T^z_{\bf j},
\end{equation}
with $E_{\rm JT} \simeq 0.9$ ($\simeq 12 $ meV for realistic 
$J_\sigma \simeq 13$ meV 
in KO$_2$~\cite{Sol08}) is enough to overcome the orbital interactions 
that follow from the spin-orbital superexchange Eq. (\ref{eq:h})
(an even smaller $E_{\rm JT} \simeq 0.04$, i.e., $\simeq 0.1$ meV suffices 
for RbO$_2$ with $J_{\sigma} \simeq 3.3$ meV~\cite{Ylv10}). Furthermore, 
recently precisely this type of robust JT-induced AO order of $T^z$ pseudospins 
was identified
[compare present Fig.~\ref{fig:1}(b) with Fig. 3(c) of Ref. \cite{Kum10}].
The estimated JT interaction at 22 meV per formula unit is
well above the minimal value of $E_{\rm JT}$. Since
the invoked mechanism relies merely on electrostatic repulsion
between electrons on alkali and oxygen atoms in bct lattice, 
we suggest that it is universal for all alkali $R$O$_2$ hyperoxides.

After explaining how the spin and concomitant orbital order can
both be stable on the frustrated bct lattice, we now present the
spin-wave dispersion Eq. (\ref{eq:omega}) in the $C$-AF phase with
$C$-AO$z$ order [Fig. \ref{fig:1}(b)] as imposed by the JT effect.
Now all magnetic couplings have a sign compatible with the
existing magnetic order, see the discussion above, and the
spin-wave dispersion (Fig. \ref{fig:4}) has no soft modes and
indicates that the ground state is stable. 
The spin-wave dispersion could be verified by future experiments.

\begin{figure}[t!]
\begin{center}
   \includegraphics[width=8.4cm]{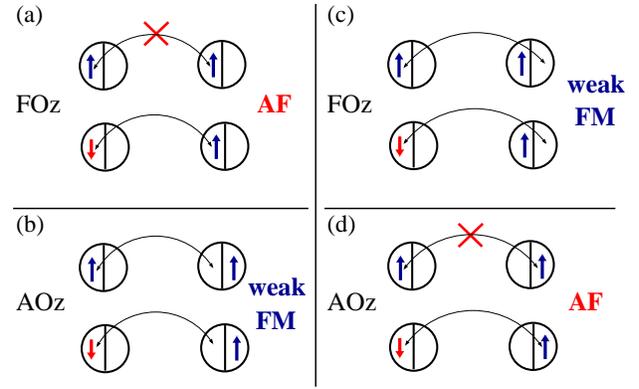}
\end{center}
\caption{(Colour on-line)
Artist's view of the FO$z$ and AO$z$ states
with: (a-b) diagonal $t_{xx}$ hopping not violating the GKR;
(c-d) interorbital $t_{xy}$ hopping fully violating the GKR.
Weak FM exchange in (b) and (c) follows from finite $\eta>0$.
}
\label{fig:5}
\end{figure}

Remarkably, the $C$-AO$z$ order supports spin-exchange constants
that are AF for FO$z$ bonds along $c_1$ as well as for AO$z$ bonds
along $c_2$ and this explains the origin of stable $C$-AF spin
order in RbO$_2$ [$\beta =1.77$ and $J_\gamma >0$ for FO and AO
state in the $c$ planes, see Fig. \ref{fig:3}(d)]. Besides, it is
very plausible that also in KO$_2$ the $C$-AF phase is stable not
because the magnitudes of the exchange constants with `wrong'
signs are small but because both exchange constants in $c$ planes
are positive; we have verified that a slightly smaller $J_H=0.4$
eV as suggested in Ref. \cite{Kum10} for KO$_2$ (and $\beta=1.9$)
yields $J_{\gamma}>0$ for FO state in the $c$ plane. Altogether,
this suggests that the classical GKR, see Figs. \ref{fig:5}(a) and
\ref{fig:5}(b), are not enforced in the $R$O$_2$ family and this
resolves the puzzle of stable $C$-AF phase.

Therefore, let us now try to understand the origin of this {\it
generalization} of the GKR by studying Hamiltonian (\ref{eq:h}) in
one of the $c$ planes only. If hoppings were almost only
interorbital, i.e., $\beta\gg 1$, the GKR would simply be inverted
and bonds with FO$z$ (AO$z$) order would drive FM (AF) couplings,
see Figs. \ref{fig:5}(c) and \ref{fig:5}(d). If both inter- and
intra-orbital hopping contribute, the two opposing tendencies
compete and the magnetic exchange is tuned by them. For a broad
range of intermediate values of $\beta$, AF interactions are
established both on the FO$z$ bonds (driven by orbital-conserving
$J_{xx}$) and on the AO$z$ bonds (driven by interorbital $\beta^2
J_{xx}$), see Fig.~\ref{fig:5}(a,d). The AF coupling dominates,
because antiferromagnetism, which is due to the Pauli principle,
is much stronger than ferromagnetism, which is caused by the
energy difference $\propto\eta$ between triplet and singlet
intermediate states of the superexchange processes.

The JT effect is crucial for this generalization of the GKR ---
without it large interorbital hopping $t_{xy}$ orders the $T^x$
pseudospin component instead of the $T^z$ component in the single
$c$ plane under consideration. It becomes then more natural to
consider the basis of $T^x$ eigenstates, being
$\{(p_x+p_y)/\sqrt{2},(p_x-p_y)/\sqrt{2}\}$. In this rotated basis, 
the full hopping term (consisting of $t_{xy}$ and $t_{xx}$) is 
diagonal, while `interorbital' hopping vanishes. Classical GKR are 
then fulfilled, and spin exchanges are 
positive (negative) for FO$x$ (AO$x$) states in the $c$ plane.

Note that such generalized GKR can arise whenever the
orbital order on a bond is not solely stabilized by the same
spin-orbital superexchange Hamiltonian that determines the spin
exchange interaction. On a geometrically frustrated lattice,
another route to this behaviour can occur when the ordered orbital
component preferred by superexchange depends on the direction
\emph{and} the relative strengths fulfill certain criteria. In the
bct case discussed here, hoppings within the $ab$ plane are
diagonal in the $\{p_x,p_y\}$ basis, while hoppings along the
$c_1$ and $c_2$ planes are diagonal in the
$\{(p_x+p_y)/\sqrt{2},(p_x-p_y)/\sqrt{2}\}$ basis. Since the $c_1$
and $c_2$ bonds frustrate each other as long as the traditional
GKR hold (see above), it follows that the $C$-AF
order can be stable on the bct lattice only if the orbital order
is the one preferred by the $ab$ plane. This appears to be
somewhat counterintuitive, as there is only one $ab$ plane and two
$c_1$ and $c_2$ planes --- it requires either a JT effect (as
here) or $ab$ hoppings that are considerably stronger than those
along $c_1$ and $c_2$. This second scenario is not 
expected for realistic parameters of $R$O$_2$ but in principle could also be
possible in a frustrated lattice. For instance, it might play a role for $d$ orbitals with
$t_{2g}$ symmetry on a triangular lattice, where hopping is
strongly anisotropic and `interorbital' along all bonds, so that
--- at least some --- orbital interactions are always
frustrated~\cite{Nor08}.

We have shown that the onset of the AF order on the frustrated 
bct lattice in $R$O$_2$ systems requires a generalization of the
well-known Goodenough-Kanamori rules, because ordered states that
obey these classical rules are destabilized by 
spin and orbital waves. A generalization arises in the presence of large
interorbital hopping whenever the orbital order enforced by
Jahn-Teller coupling is qualitatively different from the one
favoured by orbital superexchange.
We emphasize that to the best of our knowledge, there exists no
alternative explanation of the origin of the $C$-AF order in this
class of compounds on the {\it frustrated} bct lattice. For
instance, a recent study using the spin-orbital model at $\eta=0$
(see footnote \footnote{ For $\eta=0$, spin exchange vanishes in
$ab$ planes for the considered $p$ orbital order, similar to $ab$
planes in the $e_g$ compound KCuF$_3$~\cite{Fei97}.
}) starts from the assumption of $C$-AF order~\cite{Ylv10}.
Furthermore, neither the spin-orbit coupling suggested in Ref.
\cite{Sol08} to explain the high temperature behaviour of KO$_2$
nor the indirect kinetic exchange interaction can explain the
onset of the FM planes in the $R$O$_2$ hyperoxides (see footnote
\footnote{ Although a weak tendency toward FM order has been found
(within generalized gradient approximation \cite{Kov09}) in
KO$_2$, the crucial dependence on the type of tilting of O$_2$
molecules suggests that this is not a generic mechanism capable of
explaining the $C$-AF order in all alkali hyperoxides with
different types of distortions \cite{Lab79}. }).

Summarizing, alkali $R$O$_2$ hyperoxides are different from both
`plain vanilla' $p$-orbital systems in optical lattices
with effective interactions of purely electronic origin,
and from $d$-orbital compounds like the manganites. In the latter
case the superexchange and Jahn-Teller coupling support the same
orbital order and the standard Goodenough-Kanamori rules are
perfectly obeyed~\cite{Fei99}.

\acknowledgments
We thank Jeroen van den Brink for insightful discussions. 
K.W. is supported by the Alexander von 
Humboldt Foundation, M.D. by the DFG (Emmy-Noether Program).
A.M.O. acknowledges support of the Foundation for Polish Science
(FNP) and the Polish Ministry of Science and Higher Education
under Project No. N202 069639.



\begin{thebibliography}{10}

\bibitem{Lab79}  Labhart M. {\it et al.},
                   {\it Phys. Rev. B\/} \textbf{20} (1979) 53.

\bibitem{Ima98} Imada M., Fujimori A. and Tokura Y.,
                   {\it Rev. Mod. Phys.\/} \textbf{70} (1998) 1039.

\bibitem{Kug82} K.I. Kugel and D.I. Khomskii,
                   Usp. Fiz. Nauk \textbf{136}, 621 (1982)
                  [Sov. Phys. Usp. \textbf{25}, 231 (1982)].

\bibitem{Nag00} Tokura Y. and Nagaosa N.,
                   {\it Science\/} \textbf{288} (2000) 462.

\bibitem{Tok06} Tokura Y.,
                   {\it Rep. Prog. Phys.\/} \textbf{69} (2006) 797.

\bibitem{Fei97} Feiner L. F., Ole\'s A. M. and Zaanen J.,
                   {\it Phys. Rev. Lett.\/} \textbf{78} (1997) 2799;
                   {\it J. Phys.: Condens. Matter\/} \textbf{10} (1998) L555.

\bibitem{Nus04} Nussinov Z. {\it et al.},
                   {\it Europhys. Lett.\/} \textbf{67} (2004) 990.

\bibitem{Nus08} Nussinov Z. and Ortiz G.,
                   {\it Europhys. Lett.\/} \textbf{84} (2008) 36005.

\bibitem{Tro10} Trousselet F., Ole\'s A.~M. and Horsch P.,
                   {\it Europhys. Lett.\/} \textbf{91} (2010) 40005.

\bibitem{Ole06}  Ole\'s A. M. {\it et al.},
                   {\it Phys. Rev. Lett.\/} \textbf{96} (2006) 147205.

\bibitem{Zaa93} Zaanen J. and Ole\'s A. M.,
                  {\it Phys. Rev. B\/} \textbf{48} (1993) 7197.

\bibitem{Mot99} Motome Y. and Imada M.,
                  {\it Phys. Rev. B\/} \textbf{60} (1999) 7921;
                Cuoco M., Forte F. and Noce C.,
                  {\it Phys. Rev. B\/} \textbf{73} (2006) 094428.

\bibitem{vdB01} van den Brink J.,
                   {\it Phys. Rev. Lett.\/} \textbf{87} (2001) 217202.

\bibitem{Gru02} Gr\"uninger M. {\it et al.},
                   {\it Nature\/} \textbf{418} (2002) 39.

\bibitem{Zho06} Zhou J.-S. and Goodenough J. B.,
                   {\it Phys. Rev. Lett.\/} \textbf{96} (2006) 247202;
                Horsch P. {\it et al.},
                   {\it Phys. Rev. Lett.\/} \textbf{100} (2008) 167205.

\bibitem{Sol08} Solovyev I. V.,
                   {\it New J. Phys.\/} \textbf{10} (2008) 013035.

\bibitem{Kov09} Kov\'a\v{c}ik R. and Ederer C.,
                   {\it Phys. Rev. B\/} \textbf{80} (2009) 140411;
                 Kim~M. {\it et al.},
                   {\it Phys. Rev. B\/} \textbf{81} (2010) 100409.

\bibitem{Kum10} Nandy A. K. {\it et al.},
                   {\it Phys. Rev. Lett.\/} \textbf{105} (2010) 056403.

\bibitem{Ylv10} Ylvisaker E. R., Singh R. R. P. and Pickett W. E.,
                   {\it Phys. Rev. B\/} \textbf{81} (2010) 180405.

\bibitem{Mei84} Meier R. J. and Helmholdt R. B.,
                   {\it Phys. Rev. B\/} \textbf{29} (1984) 1387.

\bibitem{Lew11} Lewenstein M. and Liu W. V.,
                   {\it Nature Phys.\/} \textbf{7} (2011) 101.

\bibitem{Wir11} Wirth G., Olschlager M. and Hemmerich A.,
                   {\it Nature Phys.\/} \textbf{7} (2011) 147.

\bibitem{Wu07}  Wu C. {\it et al.}, 
                   {\it Phys. Rev. Lett.\/} \textbf{99} (2007) 070401;
                Wu C. and Das Sarma S.,
                   {\it Phys. Rev. B\/} \textbf{77} (2008) 235107.

\bibitem{Zha08} Zhao E. and Liu W. V.,
                   {\it Phys. Rev. Lett.\/} \textbf{100} (2008) 160403;
                Wu C.,
                   {\it Phys. Rev. Lett.\/} \textbf{100} (2008) 200406.

\bibitem{Goo63} Goodenough J. B.,
                   {\it Magnetism and the Chemical Bond\/}
                   (Interscience, New York, 1963);
                Kanamori J.,
                   {\it J. Phys. Chem. Solids\/} \textbf{10} (1959) 87.

\bibitem{Cha88} Chandra P. and Doucot B.,
                   {\it Phys. Rev. B\/} \textbf{38} (1988) 9335.

\bibitem{vdB99} van den Brink J. {\it et al.},
                   {\it Phys. Rev. B\/} \textbf{59} (1999) 6795.

\bibitem{Nor08} Normand B. and Ole\'{s} A. M.,
                   {\it Phys. Rev. B\/} \textbf{78} (2008) 094427;
                Chaloupka J. and Ole\'{s} A. M.,
                   {\it Phys. Rev. B\/} \textbf{83} (2011) 094406.

\bibitem{Fei99} Feiner L. F. and Ole\'s A. M.,
                   {\it Phys. Rev. B\/} \textbf{59} (1999) 3295.
\end{thebibliography}
\end{document}